\documentclass[useAMS,usenatbib]{mn2e}

\usepackage{times}
\usepackage{graphicx}

\begin{document}

 \title[Do Solar system tests permit higher dimensional general relativity?]
 {Do Solar system tests permit higher dimensional general relativity?}
 \author[F. Rahaman, Saibal Ray, M. Kalam \& M. Sarker]
 {F. Rahaman$^{1}$\thanks{E-mail: farook\_rahaman@yahoo.com},
 Saibal Ray$^{2}$\thanks{E-mail: saibal@iucaa.ernet.in}, M. Kalam$^{3}$ and M.
 Sarker$^{1}$\\
  $^{1}$Department of Mathematics, Jadavpur University, Kolkata 700 032,
 West Bengal, India\\
$^{2}$Department of Physics, Barasat Government College, North 24
Parganas, Kolkata 700 124, West Bengal, India \\ and
Inter-University Centre for Astronomy \& Astrophysics, Post Bag 4,
Ganeshkhind, Pune 411 007, India\\ $^{3}$Department of Physics,
Netaji Nagar College for Women, Regent Estate, Kolkata 700 092,
West Bengal, India}

\date{Accepted . Received ; in original form }

\pagerange{\pageref{firstpage}--\pageref{lastpage}} \pubyear{2007}

\maketitle

\begin{abstract}
We perform a survey whether higher dimensional Schwarzschild
space-time is compatible with some of the solar system phenomena.
As a test we examine five well known solar system effects, viz.,
(1) Perihelion shift, (2) Bending of light, (3) Gravitational
redshift, (4)  Gravitational time delay and (5) Motion of test
particle in the framework of general relativity with higher
dimensions. It is shown that the results related to all these
physical phenomena are mostly incompatible with the higher
dimensional version of general relativity except that of Motion of
test particle. We compare all these results with the available
data in the literature.
\end{abstract}


\section{Introduction}
Einstein's general relativity (GR) received first instant success
due to the observational confirmation of two solar system effects,
firstly, contribution to perihelion shift of $43$ arcsec per
century as curvature effect of space-time, and secondly, total
solar eclipse in the year 1919 which admits the relativistic value
$1.75$ arcsec as obtained by Einstein which is also due to the
effect of curvature. Probably these observational boldness and the
sublime structure of GR inspired \citet{Born1962} to state that,
{\it ``The theory appeared to me then, and it still does, the
greatest feat of human thinking about nature, the most amazing
combination of philosophical penetration, physical intuition, and
mathematical skill. It appealed to me like a great work of art
...''}.

The above mentioned two triumph of GR is obviously based on its
usual four-dimensional structure of space-time. This prompted
people to start thinking of the multidimensional structure of GR.
However, the extension of GR by the inclusion of dimensions beyond
four were initiated by investigators mainly in connection to the
studies of early Universe. It is commonly believed that the
four-dimensional present space-time is the compactified form of
manifold with higher dimensions (HD). This self-compactification
of multidimensions have been thought of by several researchers
\citep{Schwarz1985,Weinberg1986} in the area of grand unification
theory as well as in superstring theory.

In the Kaluza-Klein gravitational theory with higher dimensions,
 therefore, it is a common practice to show that extra dimensions
 are reducible to lower one, specially in four-dimension which
 was associated with some physical processes. Interestingly, mass have been considered as
 the fifth dimension \citep{Wesson1983,Fukui1987,Banerjee1990,Chatterjee1990,Ponce2003} in
 the case of five-dimensional Kaluza-Klein theory. \citet{Fukui1987} suggested that
 expansion of the Universe follows by the percolation of radiation
 into $4$-dimensional space-time from the fifth dimensional mass.
\citet{Ponce2003} argued that the rest mass of a particle,
perceived by an observer in four-dimension, varies as a result of
the five-dimensional motion along the extra direction and in the
presence of elctromagnetic field is totally of elctromagnetic
origin which has confirmed by \citet{Ray2006}. On the other hand,
it have been shown by many investigators
\citep{Ishihara1984,Gegenberg1985} that within the Kaluza-Klein
inflationary scenario of
 HD a contraction of the internal space causes the
 inflation of the usual space. There are cases in FRW cosmologies
 where the extra dimensions contract as a result of
 cosmological evolution \citep{Iba`nez1986}.
 In the solution to the vacuum field equations of GR in $4+1$
dimensions \citet{Chodos1980} have shown that it leads to a
cosmology which at the present epoch has $3+1$ observable
dimensions in which the Einstein-Maxwell equations are obeyed.

Under these theoretical background, therefore, now-a-days people
have started to think of the higher dimensional influence on GR,
more precisely, whether within the framework of higher dimensional
GR the same type of solar system tests would yield the same
results. Actually, it has two-fold intentions: firstly, if the
results are positive then the higher dimensional version of GR
will prove itself as an extended viable theory of gravitation, and
secondly, if negative then there is no need of higher dimensional
GR at all. Motivated by this, therefore, in a recent work
\citet{Liu2000} argued that to test the theory involving the
motion of test particles in the field of a static
spherically-symmetric mass like the Sun or the Earth would be most
straightforward. \citet{Kagramanova2006} have investigated Solar
system effects in Schwarzschild-Sitter space-time and estimated
the values for the cosmological parameter $\Lambda$. In a similar
line of thinking \citet{Iorio2005a,Iorio2005b} attempted to
investigate secular increase of the Astronomical Unit, perihelion
precessions and planetary motions as tests of the
Dvali-Gabadadze-Porrati multidimensional braneworld scenario.

In this connection it is to be noted here that investigations by
\citet{Liu2000}, along with those of \citet{Lim1995} and
\citet{Kalligas1995}, are limited to five-dimensional soliton-like
space-time only. Therefore, our present attempt is to study more
general cases under a spherically symmetric Schwarzschild-like
space-time with $N$ number of dimensions where $N=D+2$ such that
$D \geq 2$. In this context we discuss the following five cases
involved in the solar system experiments to examine the viability
of GR with HD, viz., (1) Perihelion shift (2) Bending of light (3)
Gravitational Red-shift (4) Gravitational time delay and (5)
Motion of test particle. Our present studies show that most of
these solar system phenomena do not allow dimensions beyond $4$
indicating a gross failure of GR with higher dimensional
framework.

\section{Mathematical Formulation}
 Let us consider a spherically symmetric metric which represents
 a generalized Schwarzschild space-time with higher
 dimensions \citep{Mayers1986}
\begin{equation}
ds^{2} = - f(r) dt^2 + \frac{dr^2}{f(r)} + r^2 {d {\Omega}_D}^2,
\end{equation}
where $r$ is a radial coordinate and $f$ is a function of $r$
only. The line element ${d{\Omega}_D}^2$ on the unit $D$-sphere is
given by
\begin{eqnarray}
{d{\Omega}_D}^2 = d {{{\theta}_1} ^2} + sin^2 {\theta}_1 d
{{{\theta}_2} ^2}+ sin^2 {\theta}_1 sin^2 {\theta}_2 d
{{{\theta}_3} ^2}+ ...\nonumber\\+\prod_{n=1}^{D-1} sin^2
{\theta}_n d{{\theta}_D} ^2
\end{eqnarray}
 with ${\Omega}_D=2[{\pi}^{(D+1)/2}]/[\Gamma(D+1)/2]$.
Also, according to Einstein equations we can write
$f(r)=1-\mu/r^{D-1}$ with the constant of integration $\mu=16\pi
GM/D c^2 {\Omega}_D$.

Now, in principle, in Lagrangian mechanics the trajectory of an
object is derived by finding the path which minimizes the action,
a quantity which is the integral of the Lagrangian over time. So,
in connection to the solar system problem we would like to adopt
the higher dimensional Lagrangian which can be written as
\begin{eqnarray}
L=T-V=-f{\dot t}^2 +\frac{{\dot r}^2}{f}+ r^2 \dot{{{\theta}_1}
^2}+r^2 sin^2 {\theta}_1 \dot{{\theta}_2}^2 + ...\nonumber\\ + r^2
\prod_{n=1}^{D-1} sin^2 {\theta}_n \dot{{\theta}_D} ^2.
\end{eqnarray}
 Here dot over any parameter implies differentiation with respect to the
affine parameter `s'.

 Now, if we take a cross-section by keeping fixed
${\theta}_1={\theta}_2=...={\theta}_{D-1}=\frac{\pi}{2}$, so that
${{\dot\theta}_i}=0, i=1,2,3,...,D-1$ then the Lagrangian takes
the form
\begin{equation}
L=-f{\dot t}^2 +\frac{{\dot r}^2}{f}+ r^2 \dot{{{\theta}_D} ^2}
\end{equation}
with light-like particle photon, $L=0$ and for any time-like
particle, $L=1$.

Therefore, in terms of the generalized coordinates $q_i$ and
generalized velocities $\dot q_i$, the standard Euler-Lagrange
equations are
\begin{equation}
\frac{d}{ds}\left(\frac{\partial L}{\partial \dot q_i}\right)
-\frac{\partial L}{\partial  q_i}=0.
\end{equation}
By assuming $f\dot t=E$ and $r^2 \dot{{\theta}_D}=p$, where $E$
and $p$ are the energy and momentum of the particle respectively,
such that $\dot t=E/f$ and $\dot{{\theta}_D}=p/r^2$ and hence with
these notations equation (4) becomes
\begin{equation}
L=-\frac{E^2}{f} +\frac{{\dot r}^2}{f}+ \frac{p^2}{r^2}
\end{equation}
which, after simplification, can be written in the following forms
\begin{equation}
\dot r^2=Lf + E^2 -\frac{p^2 f}{r^2}
\end{equation}
and
\begin{equation}
\frac{1}{r^4}\left(\frac{dr}{d{\theta}_D}\right)^2=\frac{Lf}{p^2}+\frac{E^2}{p^2}
-\frac{f}{r^2}.
\end{equation}
Again, by substituting ${\theta}_D=\phi$ and $r=1/U$ in equation
(8), one can write
\begin{equation}
\left(\frac{dU}{d{\phi}}\right)^2=\frac{Lf}{p^2}+\frac{E^2}{p^2}
-fU^2.
\end{equation}
Now, if we write equation (7) in the form
\begin{equation}
\left(\frac{dr}{dt}\right)^2=\frac{Lf^3}{E^2}+f^2
-\frac{p^2f^3}{E^2r^2}
\end{equation}
then one can easily observe that $\frac{dr}{dt}$ vanishes at
$r=r_0$ of the closest approach to the sun. This at once yields
the relationship between momentum and energy of the particle as
follows: $p^2/E^2 = r_0^2/f_0$, where $f_0 = f(r=r_0)$. Hence, the
equation of photon becomes
\begin{equation}
\left(\frac{dr}{dt}\right)^2=f^2-\frac{f^3r_0^2}{f_0r^2}.
\end{equation}

 Thus, the time required for light to travel from $r_0$ to $r$ can be expressed as
\begin{equation}
t(r,r_0)=\int_{r_0} ^r \frac{dr}{\left[f^2-\frac{f^3{r_0}^2}{f_0
r^2}\right]^{1/2}}.
\end{equation}

\section{Solar system tests for higher dimensional GR}
\subsection{Perihelion shift}
Following the equation (8), motion of planet in the sun's
gravitational field can be written as
\begin{equation}
\frac{1}{r^4}\left(\frac{dr}{d{\phi}}\right)^2=\frac{f}{p^2}+\frac{E^2}{p^2}
-\frac{f}{r^2}.
\end{equation}
For $r=1/U$, we have
\begin{equation}
\frac{d^2 U}{d{\phi}^2}+U=\mu(D+1)U^D+\frac{\mu}{p^2}(D-1)U^{D-2}.
\end{equation}
The solution to this equation (14) is then given by the following
cases:\\

(i): $D=2$\\  By the use of successive approximation (taking
$\mu=0$ as zeroth approximation) we get the solution to the above
equation (14) in the form
\begin{equation}
U=\frac{1}{l}(1 + e cos\phi)
\end{equation}
where $l=p^2/\frac{GM}{c^2}$. Obviously, the trajectory of test
particle, i.e., planet is elliptical (see Fig. 1).

\begin{figure*}
\begin{center}
\vspace{0.5cm}
\includegraphics[width=0.8\textwidth]{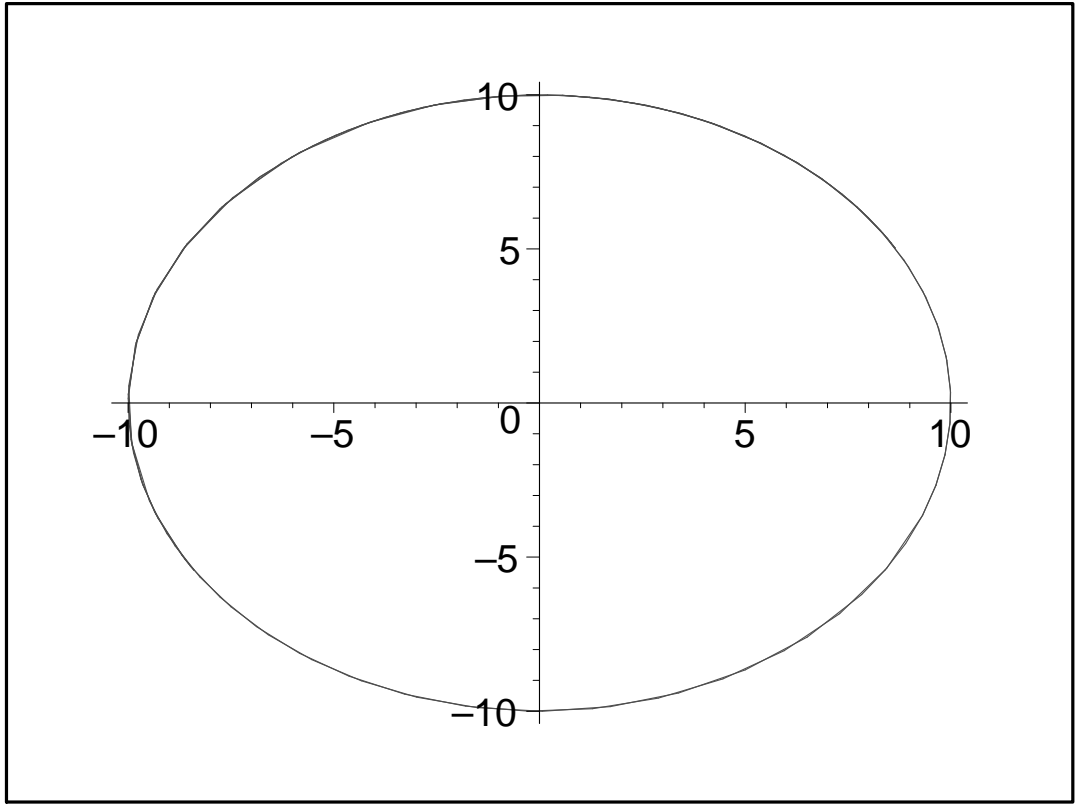}
    \caption{The trajectory of equation (15) by choosing suitably
    the parameters.}
 \label{fig:2}
\end{center}
\end{figure*}

Substituting this on the right hand side for $U$, we get
\begin{equation}
U=\frac{1}{l}[1+e cos(\phi-\omega)]
\end{equation}
with $\omega=(3GM/c^2 l)\phi$. Therefore, time period for the
planet is $T=2\pi (3GM/c^2 l)$ and the average precession can be
obtained as $n=6\pi GM/c^2 lT = 43.03$ arcsec per century (where
$l=5.53 \times 10^{12}$ cm, $GM/c^2= 1.475 \times 10^{5}$ cm and
one century$=415T$). This value of Mercury's perihelion precession
rate is very close to some of the available data which are $43.11
\pm 0.21$ and $42.98$ arcsec per century respectively as obtained
by \citet{Shapiro1976} and \citet{Liu2000}.\\

(ii): $D=3$\\ The equation of motion in this case can be written
as
\begin{equation}
\frac{d^2 U}{d{\phi}^2}+U(1-\frac{2\mu}{p^2})=4\mu U^3.
\end{equation}
The solution is given by
\begin{equation}
U=U_0 cos(\beta \psi) +U_1 cos(3\beta \psi)
\end{equation}
where $\psi = a \phi$,  $U_1 = -(2\mu/16a)U_0^2 << U_0$ and
$\beta^2 = 1-(3\mu/a)U_0^2$ with $a = 1-2\mu/p^2$. Hence, the path
is no longer elliptical (see Fig. 2).\\

\begin{figure*}
\begin{center}
\vspace{0.5cm}
 \includegraphics[width=0.8\textwidth]{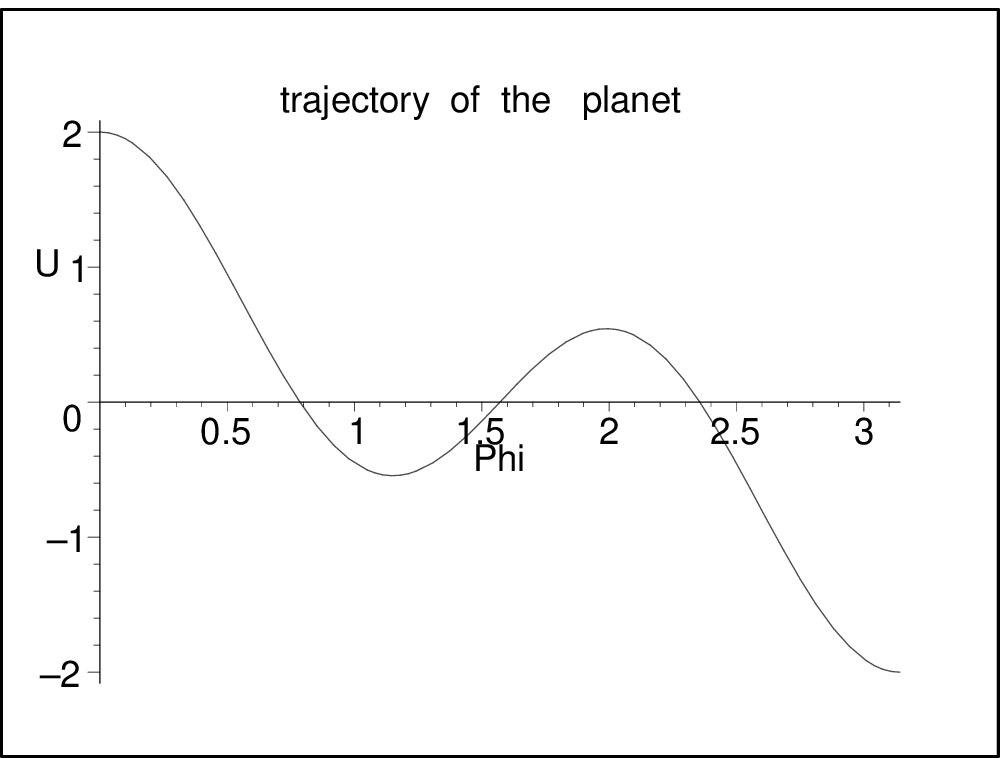}
    \caption{The plot $U$ vs. $\phi$ for $D=3$.}
 \label{fig:18}
\end{center}
\end{figure*}

(iii): $D=4$\\ The solution of the equation (14) related to motion
of planet in this case is
\begin{eqnarray}
U=\frac{cos\phi}{r_0} +\frac{5\mu}{r_0^4} \left[\frac{3}{8} -
\frac{1}{6}(2cos^2\phi-1)\right.\nonumber\\ -
\left.\frac{1}{120}(8cos^4\phi-8cos^2\phi-1)\right]\nonumber\\+\frac{3\mu}{p^2r_0^2}\left[\frac{1}{2}
- \frac{1}{3}(2cos^2\phi-1)\right].
\end{eqnarray}
Again, one can observe that the path is no longer elliptical (see
Fig. 3).

\begin{figure*}
\begin{center}
\vspace{0.5cm}
 \includegraphics[width=0.8\textwidth]{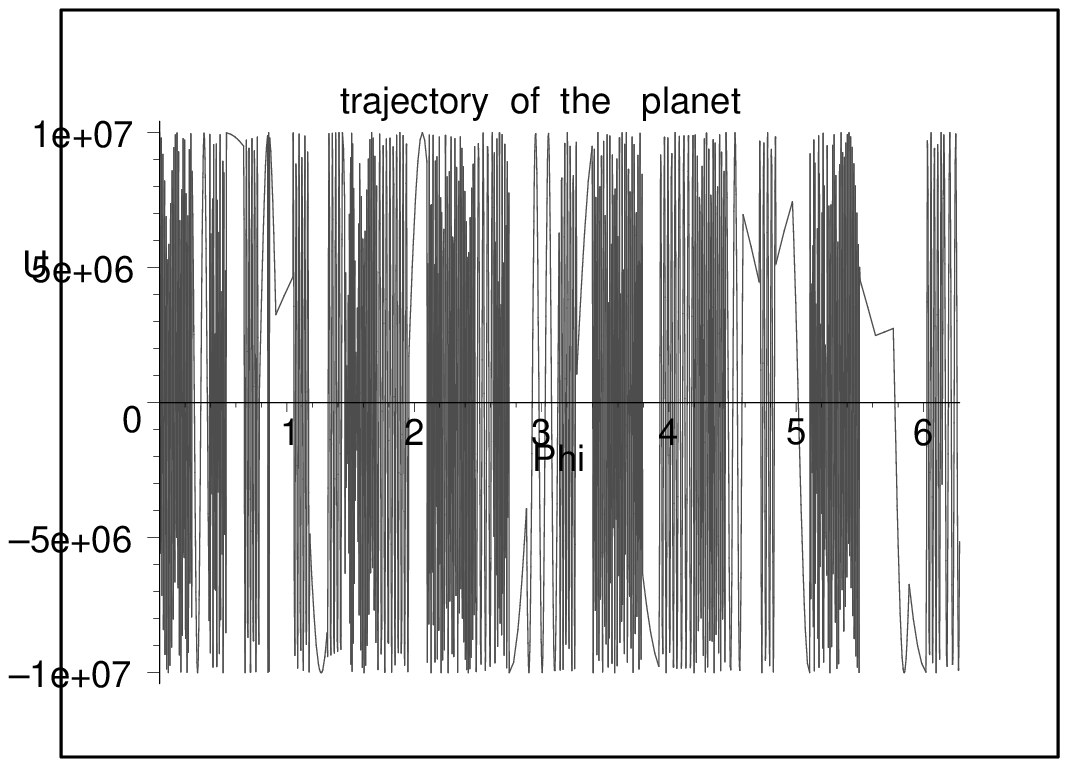}
    \caption{The plot $U$ vs. $\phi$ for $D=4$.}
  \label{fig:19}
\end{center}
\end{figure*}

\citet{Einstein1915} explained the perihelion motion of mercury
from the general theory of relativity by accounting the unsolved
amount of $\sim 43$ arcsec as due to gravitation being mediated by
the curvature of spacetime \citep{Nordtvedt2001}. Besides this
relativistic effect other effects due to classical reasons are
shown in the Table 1. Therefore, from the present investigation it
is revealed that the HD model of general relativity only admit
four-dimensional case with a precession $43.03$ arcsec per
century.

\begin{table}
\begin{center}
%\begin{minipage}{105mm}
\caption{Sources of the perihelion shift}
\label{tab 1}
\begin{tabular}{@{}llrrrrlrlr@{}}
\hline Amount  &Cause\\
                 &(arcsec per century)\\
\hline $5025.6$ &Precession of equinoxes\\ $531.4$&Gravitational
tugs of the other planets\\ $0.0254$&Oblateness of the Sun \\
$42.98 \pm 0.04$& Relativistic curvature of spacetime\\\hline
$5600.0$& Total\\\hline $5599.7$& Observed\\ \hline
\end{tabular}
%\end{minipage}
\end{center}
\end{table}

\subsection{Bending of Light}
Now we would like to observe how higher dimensional version of
general relativity do respond on the effect of light bending. Let
us, therefore, start with the equation (9) which now reads
\begin{equation}
\left(\frac{dU}{d{\phi}}\right)^2=\frac{E^2}{p^2}-U^2(1-\mu
U^{D-1}).
\end{equation}
The above equation can be written in the suitable form as
\begin{equation}
\frac{d^2 U}{d{\phi}^2}+U=\frac{\mu(D+1)}{2}U^D.
\end{equation}
Now, we solve the equation by successive approximation, starting
with the straight line (path without gravitating body) as zeroth
approximation such that $U=cos\phi/R_0$ where $\phi=0$ is the
point $P$ of nearest approach to the Sun's surface. Ideally, $R_0$
would be the solar radius.

Substituting this on the right hand side of equation (21) for $U$,
we get
\begin{equation}
\frac{d^2 U}{d{\phi}^2}+U=\frac{\mu(D+1)}{2 R_0^D}cos^D\phi.
\end{equation}
The solution of the above equation (22) is then given by for the
following cases:\\

Case I: $D=even=2n$\\ Let us consider the case when $D$ is even
and takes the value $2n$. For this particular situation the
solution to the equation (22) can be given as
\begin{eqnarray}
U=\frac{cos\phi}{R_0} +
\frac{\mu(2n+1)}{2^{2n}R_0^{2n}}\left[\frac{cos2n\phi}{-4n^2 +1}
+\frac{^{2n} C_1 cos(2n-2)\phi}{-(2n-2)^2 +1}
\right.\nonumber\\+\left.  ...
  +\frac{^{2n} C_{n-1} cos2\phi}{-2^2
+1}\right]+\frac{\mu(2n+1)^{2n}C_n }{2^{2n+1}R_0^{2n}}.
\end{eqnarray}
(i): For $n=1$\\ In this subcase
\begin{equation}
U=\frac{cos{\phi}}{R_0} + \frac{GM}{R_0^2 c^2}(2-cos^2{\phi}).
\end{equation}
For $U=0$, we get $cos\phi=-0.4244302380 \times 10^{-5}$ where the
values for the constants are taken as follows: $c=2.997925 \times
10^8$ m/sec, $G=6.67323\times 10^{-11}$ SI Unit, $M=1.9892 \times
10^{30}$ Kg and $R_0=6.95987 \times 10^8$ m. We plot $U$ vs.
$\phi$ (see Fig. 4).

\begin{figure*}
\begin{center}
\vspace{0.5cm}
 \includegraphics[width=0.8\textwidth]{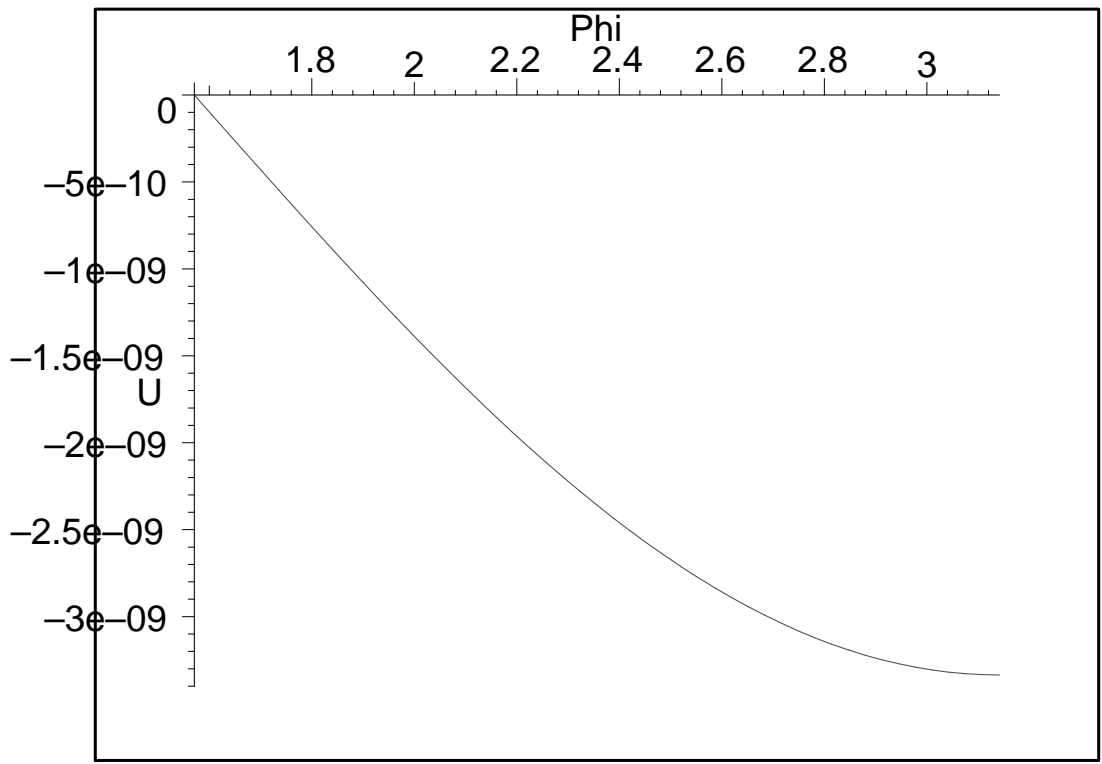}
    \caption{The plot $U$ vs. $\phi$ for $D=2$.}
  \label{fig:11}
\end{center}
\end{figure*}

Here the net deflection of the ray is given by
\begin{equation}
\Delta\phi=1.741300716 \quad arcsec.
\end{equation}
This result is in agreement with the experimental result of
observed deflection of light by the Sun (see Table 2).\\

\begin{table}
\begin{center}
%\begin{minipage}{105mm}
\caption{Deflection of Starlight During Eclipses} \label{tab 2}
\begin{tabular}{@{}llrrrrlrlr@{}}
\hline Date &Location & Deflection $(\Delta \phi)$\\
                 &&(arcsec)\\

\hline 29 May 1919   & Sobral          & $1.98 \pm 0.16$\\
                     &Principe       & $1.61 \pm 0.40$\\

21 Sep 1922    & Australia      & $1.77 \pm 0.40$\\
                               &&$1.42$ to $2.16$\\
                               &&$1.72 \pm 0.15$\\
                               &&$1.82 \pm 0.20$\\

9 May 1929    &Sumatra      &$2.24 \pm 0.10$\\

19 June 1936  &USSR          &$2.73 \pm 0.31$\\
              &Japan        &$1.28$ to $2.13$\\

20 May 1947   &Brazil          &$2.01 \pm 0.27$\\

25 Feb 1952    & Sudan          &$1.70 \pm 0.10$\\

30 June 1973 &Mauritania  &$1.66 \pm 0.19$\\ \hline
\end{tabular}
%\end{minipage}
\end{center}
\end{table}

(ii): $n=2$\\ Here
\begin{equation}
U=\frac{cos\phi}{R_0} - \frac{\mu}{6R_0^4}cos^4\phi -
\frac{2\mu}{3R_0^4}cos^2\phi +\frac{4\mu}{3R_0^4}.
\end{equation}
For $U=0$, we get $cos\phi=$imaginary. We plot $U$ vs. $\phi$ (see
Fig. 5).\\

\begin{figure*}
\begin{center}
\vspace{0.5cm}
 \includegraphics[width=0.8\textwidth]{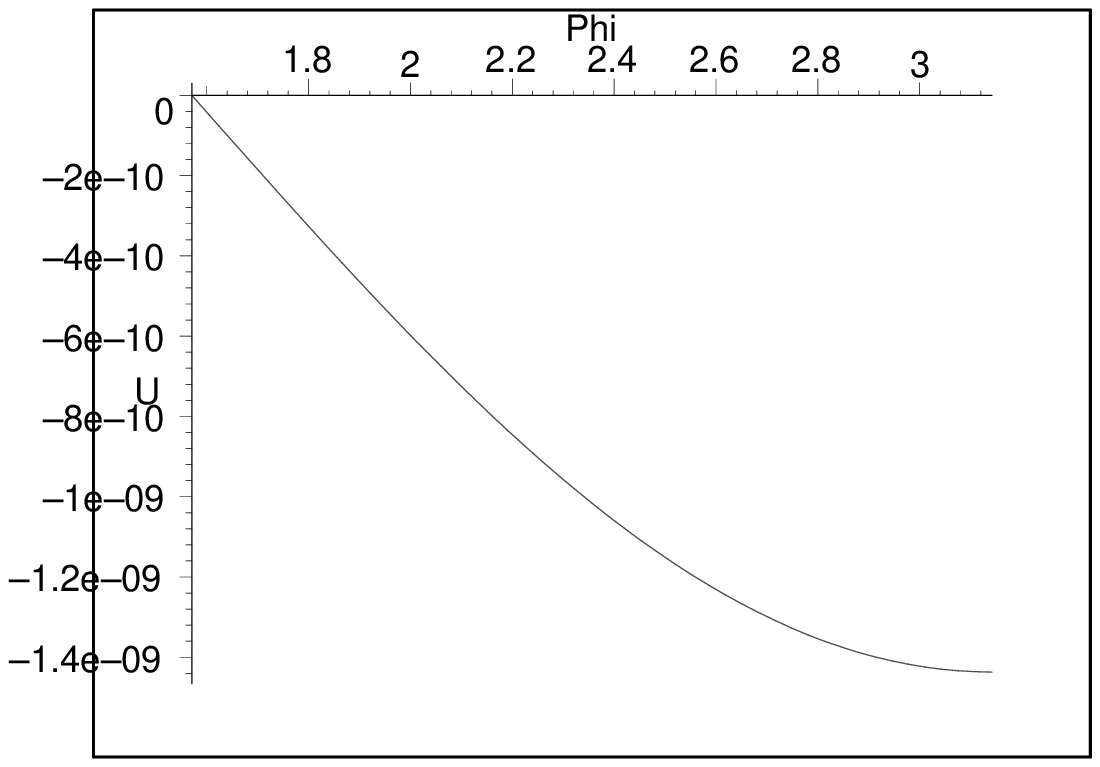}
    \caption{The plot $U$ vs. $\phi$ for $D=4$.}
  \label{fig:12}
\end{center}
\end{figure*}

(iii): $n=3$\\ For the value $n=3$, we get
\begin{eqnarray} U=\frac{cos\phi}{R_0}
-\frac{7\mu}{64R_0^6}\left[\frac{1}{35}(32cos^6\phi-48cos^4\phi+8cos^2\phi
-1) \right. \nonumber\\ + \left.
\frac{2}{5}(8cos^4\phi-8cos^2\phi+1) +5(2cos^2\phi
-1)\right]+\frac{35\mu}{32R_0^6}\end{eqnarray}
 Here $U=0$ yields $cos\phi=-0.2255226204 \times 10^{-41}$ so that
 $\Delta\phi=-0.948825313 \times 10^{-2}$ arcsec. This value of
$\Delta\phi$ expresses the angle of surplus rather than angle of
deficit \citep{Dyer1995,Rahaman2005a}. We plot $U$ vs. $\phi$ (see
Fig. 6).\\

\begin{figure*}
\begin{center}
\vspace{0.5cm}
\includegraphics[width=0.8\textwidth]{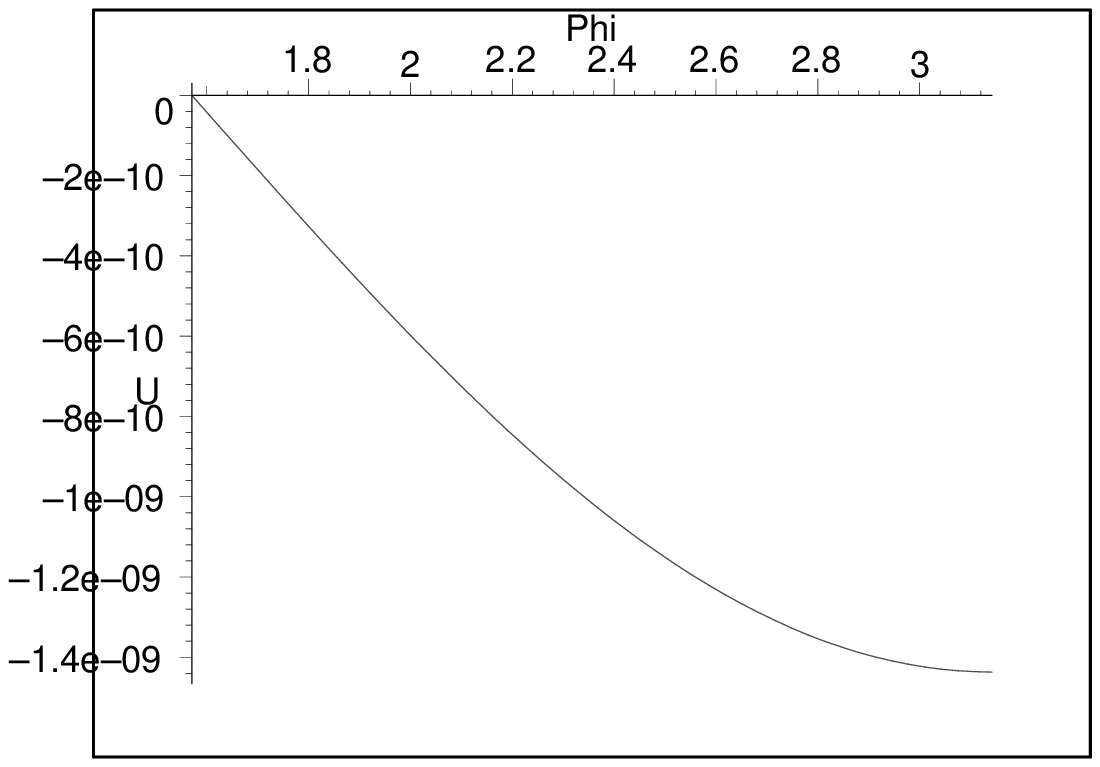}
    \caption{The plot $U$ vs. $\phi$ for $D=6$.}
  \label{fig:13}
\end{center}
\end{figure*}

Case II: $D = odd = 2n-1$
\begin{eqnarray}
U=\frac{cos\phi}{R_0} + \frac{\mu
n}{2^{2n-2}R_0^{2n-1}}\left[\frac{cos(2n-1)\phi}{-(2n-1)^2+1}
\right. \nonumber\\  + \left.\frac{^{2n-1}C_1
cos(2n-3)\phi}{-(2n-3)^2 +1}+..... \right] + \mu
n\frac{^{2n-1}C_{n-1}}{2^{2n-2}R_0^{2n-1}}\frac{\phi sin\phi}{2}
\end{eqnarray}

(i): $n=2$\\ In this subcase
\begin{equation}
U=\frac{cos\phi}{R_0} +\frac{2\mu}{R_0^3}
\left[\frac{3\phi}{8}sin\phi - \frac{1}{32}cos3\phi\right].
\end{equation}
For $U=0$, we get $cos\phi=-0.2255226204 \times 10^{-41}$. We plot
$U$ vs. $\phi$ (see Fig. 7). Here the net deflection of the ray is
given by
\begin{equation}
\Delta\phi=-0.948825313 \times 10^{-2} \quad arcsec,
\end{equation}
which is nothing but angle of surplus.\\

\begin{figure*}
\begin{center}
\vspace{0.5cm}
 \includegraphics[width=0.8\textwidth]{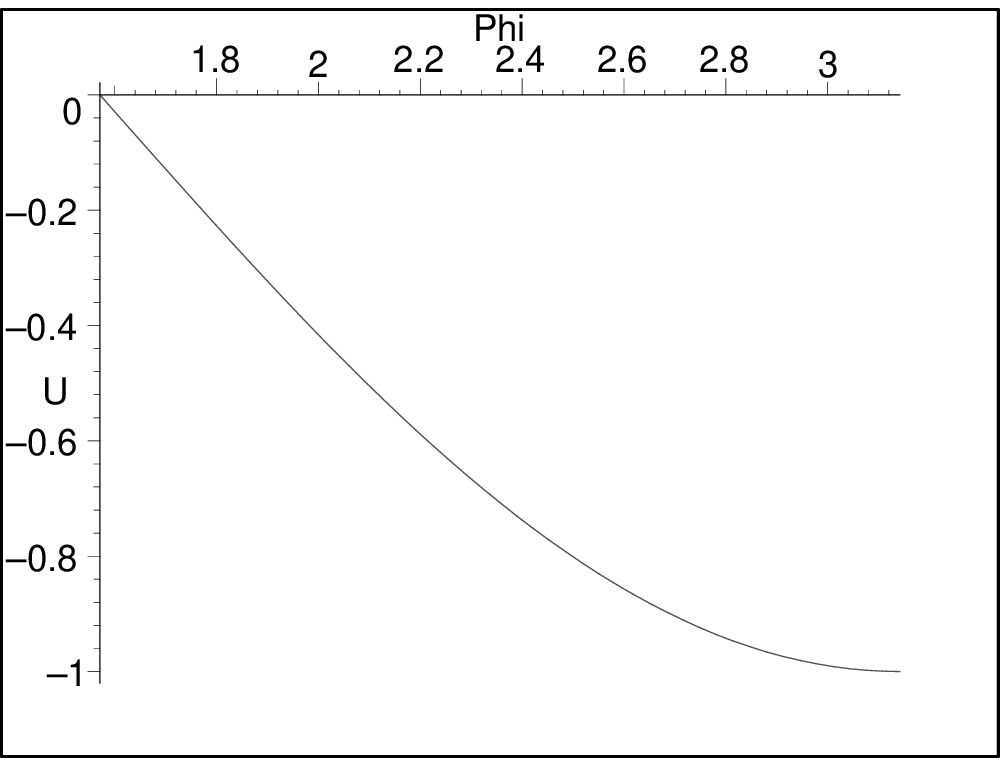}
    \caption{The plot $U$ vs. $\phi$ for $D=3$.}
  \label{fig:14}
\end{center}
\end{figure*}

(ii): $n=3$\\ In this subcase
\begin{equation}
U=\frac{cos\phi}{R_0} +\frac{3\mu}{R_0^5}
\left[\frac{5\phi}{16}sin\phi -
\frac{5}{128}cos3\phi-\frac{1}{384}cos5\phi\right].
\end{equation}
For $U=0$, we get $cos\phi=-0.2255226204 \times 10^{-41}$. We plot
$U$ vs. $\phi$ (see Fig. 8). Here the net deflection of the ray is
given by
\begin{equation}
\Delta\phi=-0.948825313 \times 10^{-2} \quad arcsec,
\end{equation}
which is again angle of surplus.

\begin{figure*}
\begin{center}
\vspace{0.5cm}
 \includegraphics[width=0.8\textwidth]{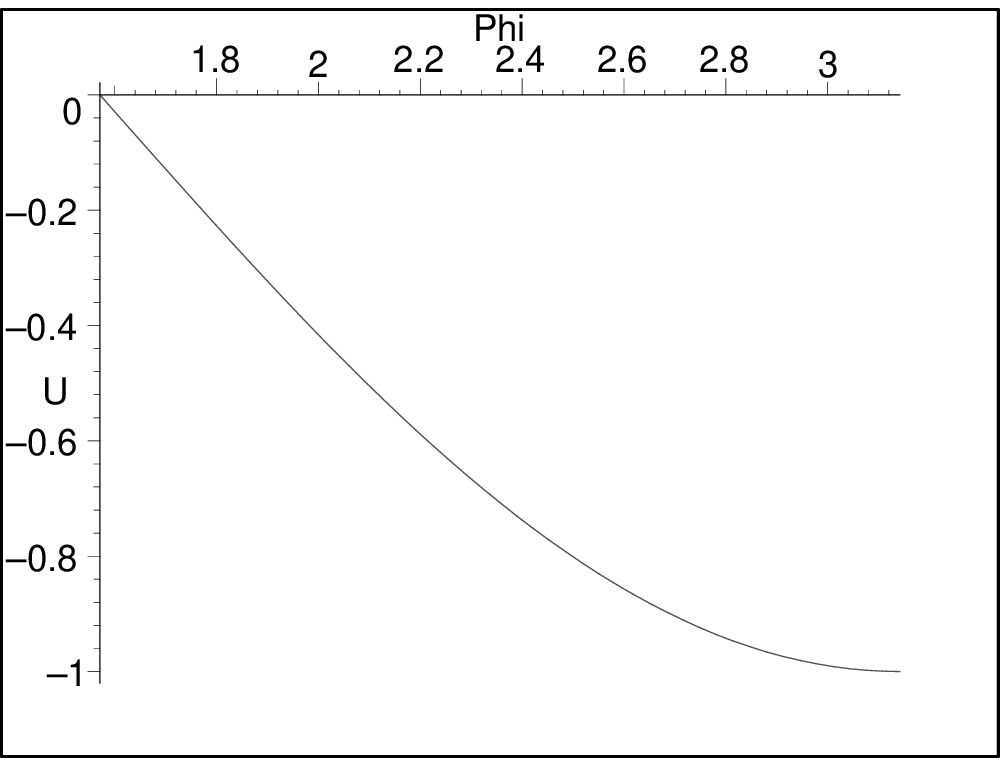}
    \caption{The plot $U$ vs. $\phi$ for $D=5$.}
  \label{fig:15}
\end{center}
\end{figure*}

Thus, we observe that when $D=2$, viz., the total dimensions,
$N=D+2=4$ then only the result does agree with the observational
data and $D>2$, i.e., $N>4$ is not compatible with solar system
(see Table 3). It can be noted,  from the Figs. 4 - 8, that all
the trajectories of the light rays almost same for all $D$ due to
the factor $cos\phi/R_0$ which is the dominating one. The
trajectory of light will show different graph for large $D$.

\begin{table}
\begin{center}
%\begin{minipage}{105mm}
\caption{Deflection of Starlight for HD-Models} \label{tab 3}
\begin{tabular}{@{}llrrrrlrlr@{}}
\hline Dimensions $(N)$  &Deflection $(\Delta \phi)$\\
                 &(arcsec)\\
\hline 4          & $1.741300716$\\

5         &  $-0.948825313 \times 10^{-2}$\\

6         & ...\\

7         &$-0.948825313 \times 10^{-2}$\\

8   &$-0.948825313 \times 10^{-2}$\\ \hline
\end{tabular}
%\end{minipage}
\end{center}
\end{table}

Historically, it is important to note that on the basis of his
`corpuscular' theory including laws of mechanics and gravitation,
\citet{Newton1704} raised the pertinent issue that ``Do not Bodies
act upon Light at a distance, and by their action bend its Rays,
and is not this action strongest at the least distance?'' He
calculated the amount of bending of light rays for Sun as
$2m/r_0$. For $m=1475$ meters, in the gravitational units, and
$r_0=6.95 \times 10^8$ meters this equals $0.875$ arcsec. However,
though prediction of bending by \citet{Einstein1911} was at first
identical to that of Newton but later on he (1915) got the angular
deflection of light as {\it twice} the size he predicted earlier
which caused due to the general relativistic effect of the curved
space-time. In 1919 scientific expeditions performed at Sobral in
South America and Principe in West Africa by the leadership of
Eddington. The reported observational results of angular
deflections due to the solar eclipse were $1.98 \pm 0.16$ and
$1.61 \pm 0.40$ arcsec, respectively. The mean of these two data
was taken as confirmation of Einstein's prediction of $1.75$
arcsec (see Table 2). However, the experiments of Eddington and
his co-workers had only $30$ percent accuracy where the results
were scattered between one half and twice the Einstein value
\citep{Will2001}. An analysis of large amount of Very Long
Baseline Interferometry (VLBI) observations has shown that the
ratio of the actual observed deflections to the deflections
predicted by general relativity is very close to unity (e.g.,
$0.9996 \pm 0.0017$ \citep{Lebach1995}, $0.99994 \pm 0.00031$
\citep{Eubanks1999}, $0.99992 \pm 0.00023$ \citep{Shapiro2004}).

\subsection{Gravitational Redshift}
GR predicted that the frequency of the light would be affected due
to gravitational field and is observable as a shift of spectral
lines towards the red end of the spectrum. Pound-Rebka-Snider
\citep{Pound1959,Pound1960,Pound1964} confirmed this effect
through their precision test, sometimes known as Harvard Tower
Experiment. In their first test they measured the redshift
experienced by a 14.4 Kev $\gamma$-rays from the decay of
$Fe^{57}$ for a height of 22.5 meter tower and found $z=2.57 \pm
0.26 \times 10^{-15}$.

Now, as usual, gravitational redshift for the solar system can be
defined as
\begin{equation}
z=\frac{\Delta\gamma}{\gamma}=\left[\frac{g_{tt}(R^{\star})}{g_{tt}(R)}\right]^{1/2}
-1
\end{equation}
where $R$ is the radius of the sun and $R^{\star}$ is the radius
of the earth's orbit around sun.

Therefore, in view of the given HD line element (1), the metric
tensors involved in the above equation (33) reduce to
\begin{eqnarray}
\left[\frac{g_{tt}(R^{\star})}{g_{tt}(R)}\right]^{1/2}=
\left[\frac{1-\frac{\mu}{R^{{\star}{D-1}}}}{1-\frac{\mu}{R^{D-1}}}\right]^{1/2}\nonumber
\\ \cong \left[1+\frac{\mu}{2R^{D-1}}-
\frac{\mu}{2R^{{\star}{D-1}}}\right].
\end{eqnarray}
By substituting the expressions of equation (34) in equation (33)
for the assumption $R^{\star}>>\mu$, we get
\begin{equation}
z=\frac{\Delta \gamma}{\gamma}=\frac{\mu}{2R^{D-1}}.
\end{equation}
Thus, in the Sun-Earth system we observe that for the usual
4-dimensional case ($D=2$), gravitational redshift becomes $z \sim
2.12 \times 10^{-6}=z_2$ (say). Therefore, for $D>2$, $z<z_2$
which indicates that as dimension increases the redshift gradually
decreases (see Fig. 9). It can be also observed that redshift
gradually increases with mass (since $z \propto \mu=16\pi GM/D c^2
{\Omega}_D$, when radial distance and dimensions remain fixed in
equation (35)). Thus, it seems that dimension acts as inversely
proportional to mass  of the gravitating  body.

\begin{figure*}
\begin{center}
\vspace{0.5cm}
 \includegraphics[width=0.8\textwidth]{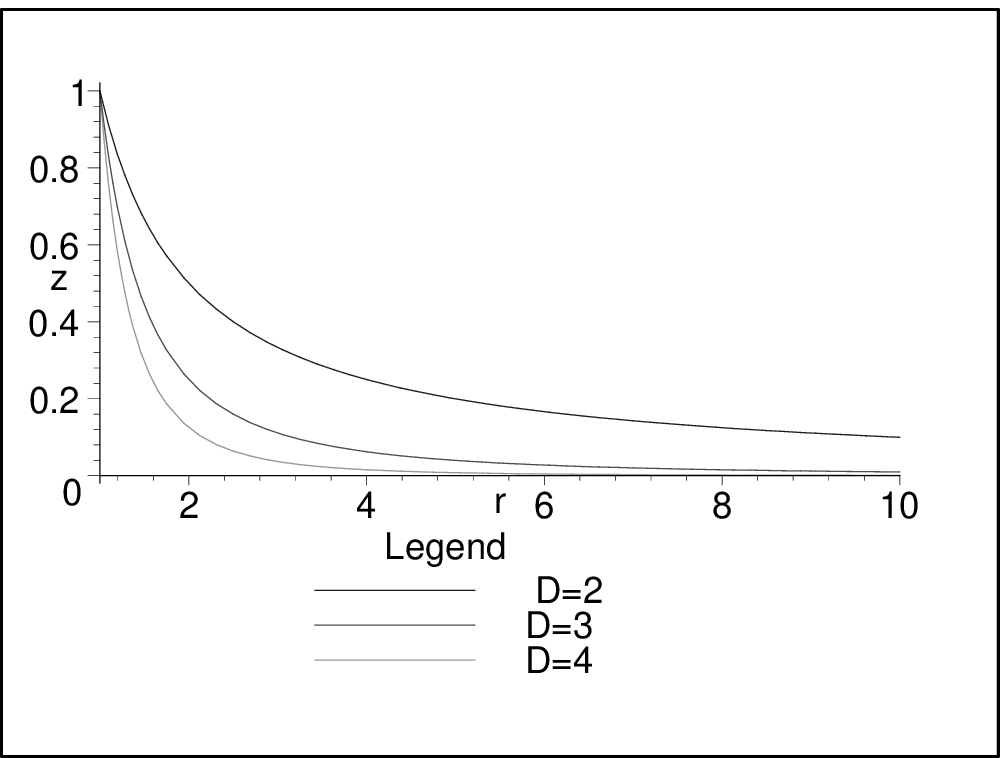}
    \caption{The plot Redshift (z) vs. $r$ for different dimensions.}
  \label{fig:redshift}
\end{center}
\end{figure*}

\subsection{Gravitational Time Delay}
Gravitational time delay, also known as Shapiro time delay which
was reported by \citet{Shapiro1964} is basically the effect of
radar signals passing near a massive object take slightly longer
time for a round trip as measured by the observer than it would be
in the absence of the object there. To proceed on towards the
`Fourth Test of General Relativity' let us consider the equation
(12) in the form
\begin{equation}
t(r,r_0)=\int_{r_0} ^r
\frac{dr}{\left(1-\frac{\mu}{r^{D-1}}\right)\left[1-\frac{1-\frac{\mu}{r^{D-1}}}{1-\frac{\mu}{r_0^{D-1}}}
\left(\frac{r_0}{r}\right)^2\right]^{1/2}}
\end{equation}
which, after simplification, yields
\begin{eqnarray}
t(r,r_0)=\int_{r_0} ^r (1-\frac{r_0^2}{r^2})^{-1/2} \times
\nonumber \\ ~~~~~~~~~~~~~~~~~~~\left[1+\frac{\mu}{r^{D-1}}
+\frac{\frac{\mu}{2}(r_0^{D-1}-r^{D-1})}{(r_0^2-r^2)r^{D-1}r_0^{D-3}}\right]dr.
\end{eqnarray}
Hence, transit time of the light ray from Mercury to Earth can be
given by
\begin{eqnarray}
t=\int_{r_0}^{r_1} (1-\frac{r_0^2}{r^2})^{-1/2}
\left[1+\frac{\mu}{r^{D-1}}
+\frac{\frac{\mu}{2}(r_0^{D-1}-r^{D-1})}{(r_0^2-r^2)r^{D-1}r_0^{D-3}}\right]dr\nonumber\\
 +\int_{r_0}^{r_2}
(1-\frac{r_0^2}{r^2})^{-1/2}\left[1+\frac{\mu}{r^{D-1}}
+\frac{\frac{\mu}{2}(r_0^{D-1}-r^{D-1})}{(r_0^2-r^2)r^{D-1}r_0^{D-3}}\right]dr.
\end{eqnarray}

In the absence of the gravitational field (viz., $\mu=0$) one can
get
\begin{equation}
t_0=\int_{r_0}^{r_1} (1-\frac{r_0^2}{r^2})^{-1/2}dr
+\int_{r_0}^{r_2} (1-\frac{r_0^2}{r^2})^{-1/2}dr.
\end{equation}
Hence, time delay for a round trip is
\begin{eqnarray} {\Delta}t=2(t-t_0)\nonumber\\=2\int_{r_0}^{r_1}
(1-\frac{r_0^2}{r^2})^{-1/2}[\frac{\mu}{r^{D-1}}
+\frac{\frac{\mu}{2}(r_0^{D-1}-r^{D-1})}{(r_0^2-r^2)r^{D-1}r_0^{D-3}}]dr\nonumber\\
 +2\int_{r_0}^{r_2} (1-\frac{r_0^2}{r^2})^{-1/2}\left[\frac{\mu}{r^{D-1}}
+\frac{\frac{\mu}{2}(r_0^{D-1}-r^{D-1})}{(r_0^2-r^2)r^{D-1}r_0^{D-3}}\right]dr.
\end{eqnarray}

Let us consider
\begin{equation}
I=\int_{r_0}^{r_1}
(1-\frac{r_0^2}{r^2})^{-1/2}\left[\frac{\mu}{r^{D-1}}
+\frac{\frac{\mu}{2}(r_0^{D-1}-r^{D-1})}{(r_0^2-r^2)r^{D-1}r_0^{D-3}}\right]dr.
\end{equation}
The solution to this equation (41) is then given by for the
following cases:\\

(i): $D=2$\\ In this case the integral in the equation (41)
becomes
\begin{equation}
I=\mu ln \left[r+\sqrt{r^2-r_0^2}\right]_{r_0}^r
+\frac{\mu}{2}\left[\sqrt\frac{r-r_0}{r+r_0}\right]_{r_0}^r
\end{equation}
so that the time delay for a round trip can be given as
\begin{eqnarray}
{\Delta}t=\frac{4GM}{c^2}
ln\left[\frac{(r_1+\sqrt{r_1^2-r_0^2})(r_2+\sqrt{r_2^2-r_0^2})}{r_0^2}\right]+\nonumber\\
\frac{2GM}{c^2}
\left[\sqrt\frac{r_1-r_0}{r_1+r_0}+\sqrt\frac{r_2-r_0}{r_2+r_0}\right].
\end{eqnarray}
If, however, $r_0<<r_1$ and $r_0<<r_2$, then
\begin{equation}
{\Delta}t=\frac{4GM}{c^2} \left[1 + ln\frac{4r_1
r_2}{r_0^2}\right].
\end{equation}
The above expression for radar echo delay is in accordance with
the standard literature \citep{Weinberg2004} when the
Schwarzschild space-time is of usual four-dimensional entity and
provides an amount $240~\mu$sec as the maximum excess time delay
for the Earth-Mercury system.\\

(ii): $D=3$\\ Here
\begin{equation}
I=\frac{3\mu}{2r_0}sec^{-1}\left(\frac{r}{r_0}\right).
\end{equation}
Hence, the time delay in this case becomes
\begin{equation}
{\Delta}t=\frac{3\mu}{r_0}\left[sec^{-1}
\left(\frac{r_1}{r_0}\right) + sec^{-1}
\left(\frac{r_2}{r_0}\right)\right].
\end{equation}
Let us consider that either $x=(r_1/r_0)>>1$ or $x=(r_2/r_0)>>1$
so that, after neglecting the higher order terms like $1/x^3$,
$1/x^5$, ... etc. we get
\begin{equation}
sec^{-1}x \cong \frac{\pi}{2} -\frac{1}{x}
\end{equation}
so that
\begin{equation}
{\Delta}t \cong \frac{4GM}{c^2} \left[\frac{1}{r_0}
-\frac{1}{\pi}\left(\frac{1}{r_1}+\frac{1}{r_2}\right)\right].
\end{equation}

(iii): $D=4$\\ For this case we have
\begin{equation}
I=\frac{3GM}{2\pi c^2}\left[\frac{\sqrt{r^2-r_0^2}}{r{r_0}^2}
+\frac{1}{r_0^2} \sqrt\frac{r-r_0}{r+r_0}\right]_{r_0}^r.
\end{equation}
Therefore, the expression for time delay becomes
\begin{eqnarray}
{\Delta}t=\frac{3GM}{\pi
c^2r_0^2}\left[\frac{\sqrt{r_1^2-r_0^2}}{r_1}+\frac{\sqrt{r_2^2-r_0^2}}{r_2}\right.\nonumber\\
+ \left.\sqrt\frac{r_1-r_0}{r_1+r_0}+
\sqrt\frac{r_2-r_0}{r_2+r_0}\right].
\end{eqnarray}
Thus, from the above case studies one can observe that the maximum
time delay will occur when $D=2$, i.e., for the usual
$4$-dimensional Schwarzschild space-time. Time delay decreases due
to increase dimensions. This can be shown easily by assuming
$r_0<<r_1$ and $r_0<<r_2$.

\subsection{Motion of Test Particle}
Let us consider a test particle having mass $m$ which is moving in
the gravitational field of a $D+2$-dimensional spacetime described
by the metric (1). So, the Hamilton-Jacobi (HJ) equation for the
test particle is \citep{Chakraborty1996,Biswas1996,Rahaman2005b}
\begin{equation}
   g^{ik}\frac{\partial S}{\partial x^i} \frac{\partial S}{\partial
   x^k}+ m^2 = 0
    \end{equation}
   where $g_{ik}$ are the classical background  field and $S$ is the
   Hamilton's characteristic function. For the metric (1) the explicit
   form of HJ equation (51) now takes the form as
\begin{eqnarray}
  - \frac{1}{f}\left(\frac{\partial S}{\partial t}\right)^2 + f\left(\frac{\partial S}{\partial
  r}\right)^2+ \frac{1}{r^2}\left[\left(\frac{\partial S}{\partial x_1}\right)^2
  + \left(\frac{\partial S}{\partial x_2}\right)^2 \right.\nonumber\\
  + \left. ... + \left(\frac{\partial S}{\partial x_{D-2}}\right)^2\right] +  m^2 = 0
   \end{eqnarray}
 where $x_1, x_2 ,.......,x_{D - 2}$ are the independent coordinates
 on the surface of the unit $(D - 2)$ sphere such that
\begin{eqnarray} d \Omega_{D-2}^2 = dx_1^2 + dx_2^2 +
.............+dx_{D-2}^2\nonumber\\ \equiv d\theta_1^2 +\sin ^2
\theta_1 d\theta_2^2+ ............... + \nonumber\\ \sin ^2
\theta_1\sin ^2 \theta_2.............\sin ^2
\theta_{D-3}d\theta_{D-2}^2
\end{eqnarray}
and $f$ is given by $f(r)=1-\mu/r^{D-1}$ as introduced earlier.

  In order to solve the above partial differential equation (52), let us choose the HJ function $S$ as
  \begin{equation}
       S = - Et + S_1(r) + p_1.x_1 + p_2.x_2 +........... + p_{D - 2}.x_{D -
       2}
  \end{equation}
 where $E$ is identified as the energy of the particle and $p_1, p_2 ,......, p_{D - 2}$
 are the momenta of the particle along different axes on the $( D - 2 )$
 sphere with the resulting momentum of the particle, $p = \sqrt{ p_1^2 + p_2^2 + ...... + p_{D - 2}^2}$.

  Now, substitution of the {\it ansatz} (54) in equation (52) provides the
  following expression for the unknown function  $S_1$ which is
\begin{equation}
         S_1(r)  =  \epsilon \int \sqrt{\frac{E^2}{f^2} - \frac{m^2}{f} - \frac{p^2}{r^2 f}} dr
 \end{equation}
with $\epsilon = \pm  1$, where the sign changes whenever $r$
passes through a zero of the integral (55).

To determine the trajectory of the particle following HJ method,
let us consider that  $\frac{\partial S}{\partial E} = constant$
and $\frac{\partial S}{\partial p_i} = constant$ [$i =
1,2,........, (D - 2)$]. Here we have chosen the constants to be
zero without any loss of generality.

  Therefore, based on the above assumptions one obtains the following two integrals
\begin{equation}
         t = \epsilon \int \frac{\frac{E}{f^2}}{\sqrt{\frac{E^2}{f^2} - \frac{m^2}{f} - \frac{p^2}{r^2 f}}} dr,
\end{equation}
\begin{equation}
         x_i = \epsilon \int \frac{(\frac{p_i}{r^2f})}{\sqrt{\frac{E^2}{f^2} - \frac{m^2}{f} - \frac{p^2}{r^2 f}}} dr.
 \end{equation}
 The radial velocity of the particle is then given, from equation (56), by
\begin{equation}
         \frac{dr}{dt} =  \frac{\sqrt{\frac{E^2}{f^2} - \frac{m^2}{f} - \frac{p^2}{r^2 f}}}{\frac{E}{f^2}}.
\end{equation}
 The turning points of the trajectory can be characterized by $\frac{dr}{dt} = 0$ and
 as a consequence the potential curve becomes
\begin{equation}
\frac{E}{m}= \sqrt f \left[1+\frac{p^2}{m^2 r^2}
\right]^{1/2}\equiv V(r)
\end{equation}
so that one can write the effective potential, $V(r)$, in the form
\begin{equation}
V^2=\left(1-\frac{\mu}{r^{D-1}}\right) \left(1+\frac{p^2}{m^2
r^2}\right)
\end{equation}
Now, in a stationary system of energy $E$, the effective potential
$V$ must have an extremal value. Therefore, the condition to be
imposed on the value of $r$ for which energy attains its extremal
one can be given by $\frac{dV}{dr}=0$ so that
\begin{equation}
2p^2 r^{D-1}-\mu (D-1)m^2 r^2-p^2 \mu(D+1)=0.
\end{equation}
It has at least one positive root the last term being negative
(for $D>3$). Thus, particles can be trapped by gravitational field
of higher dimensional Schwarzschild space-time and hence the
gravitational field is attractive in nature. For $D=2$ and $D=3$,
we have some restrictions to get bound orbit as $p^2>3{\mu}^2 m^2$
and $p^2>\mu m^2$ respectively. The plot $V^2$ vs. $r$ for $D=2$
and $D=3$ have been provided in Figs. 10 and 11 respectively.\\

\begin{figure*}
\begin{center}
\vspace{0.5cm}
\includegraphics[width=0.8\textwidth]{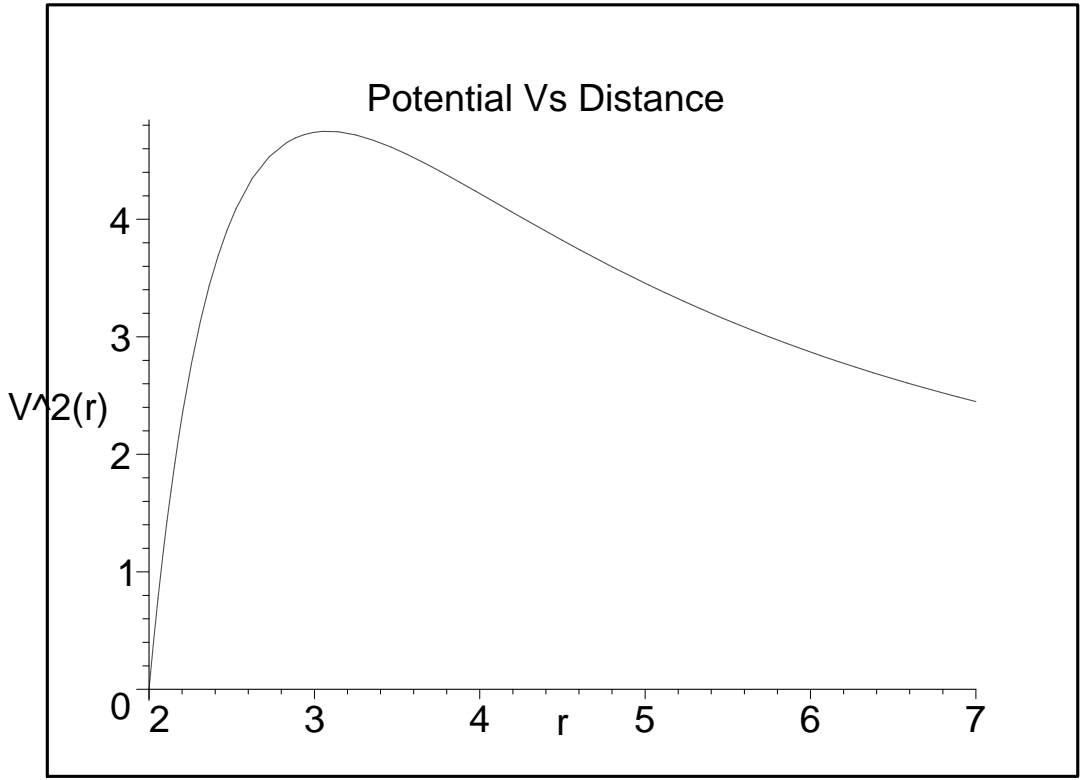}
    \caption{The plot $V^2$ vs. $r$ for $D=2$.}
 \label{fig:16}
\end{center}
\end{figure*}

\begin{figure*}
\begin{center}
\vspace{0.5cm}
\includegraphics[width=0.8\textwidth]{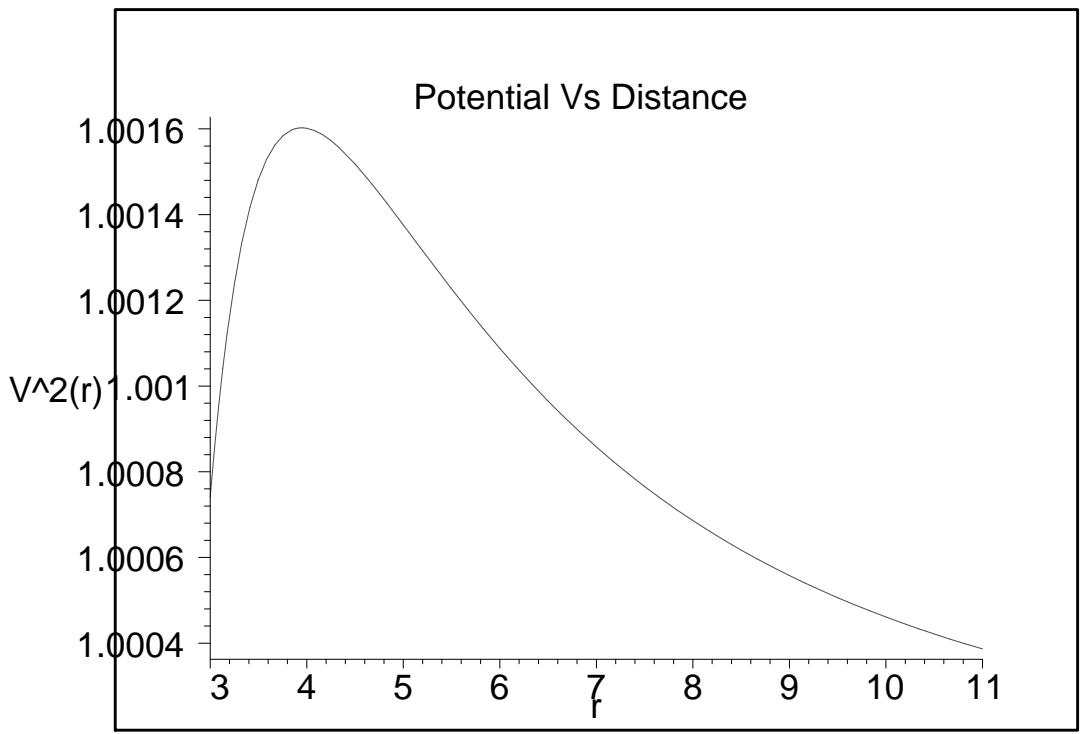}
    \caption{The plot $V^2$ vs. $r$ for $D=3$.}
 \label{fig:17}
\end{center}
\end{figure*}

\section{Conclusions}
Our analytically performed solar system tests for GR with HD can
be summarized as follows -

 1. Perihelion shift: In $4D$ our result exactly coincides with that of
 Einstein's predicted value with an elliptical path followed by the
 planet Mercury. As we go increase on dimensions the paths rapidly
 become irregular in shapes and hence HD do not work at all.

2. Bending light: Here also our theoretical result is in good
agreement with the experimental result $1.741300716$ arcsec which
become enormously different with an angle of surplus value
$-0.00948825313$ arcsec $D>2$.

3. Gravitational redshift: We observe that in the $4$-dimensional
case gravitational redshift becomes $z \sim 2.12 \times 10^{-6}$
in the Sun-Earth system. However, for $D>2$ redshift gradually
decreases with the increase of dimensions such that dimension acts
as inversely proportional to mass of the gravitating body. It can
also be observed that for constant radial distant and dimension
the redshift gradually increases with the mass of the planets.

4. Gravitational time delay: It is seen from the present
investigation that radar echo delay is as usual in the case of
$4D$ and decreases with increase of dimensions.

5. Motion of a test particle: Here the observation is that
particles can be trapped by gravitational field of higher
dimensions and hence the gravitational field is attractive in
nature (with the restrictions to get bound orbit as $p^2>3{\mu}^2
m^2$ and $p^2>\mu m^2$ for $D=2$ and $D=3$ respectively).
Therefore, this is the only case under our study which is fairly
compatible with the HD version of GR.

In a nutshell, our overall observation regarding HD realm of GR
is, in general, similar to that of \citet{Liu2000} which is as
follows: ``... the existence of small but potentially measurable
departures from the standard $4D$ Einstein predictions''. However,
in some of our HD cases invoke the word `drastic' in place of
`small' one!\\

\section*{Acknowledgments} SR is thankful to the authority of
Inter-University Centre for Astronomy and Astrophysics, Pune,
India for providing him Associateship programme under which a part
of this work was carried out.


\begin{thebibliography}{99}
\bibitem[\protect\citeauthoryear{Banerjee, Bhui \& Chatterjee}{1990}]{Banerjee1990} Banerjee A., Bhui B. K.
and Chatterjee S., 1990, Astron. Astrophys. {\bf 232} 305.

\bibitem[\protect\citeauthoryear{Born}{1962}]{Born1962} Born Max, 1962, Einstein's Theory of Relativity, (Dover).

\bibitem[\protect\citeauthoryear{Chakraborty}{1996}]{Chakraborty1996} Chakraborty S., 1996,
Gen. Rel. Grav. {\bf 28} 1115.

\bibitem[\protect\citeauthoryear{Chakraborty \& Biswas}{1996}]{Biswas1996} Chakraborty S. and Biswas L., 1996,
Class. Quan. Grav. {\bf 13} 3253.

\bibitem[\protect\citeauthoryear{Chatterjee \& Bhui}{1990}]{Chatterjee1990} Chatterjee S. and Bhui B. K., 1990,
Astrophys. Space Sci. {\bf 167} 61.

\bibitem[\protect\citeauthoryear{Chodos \& Detweiler}{1980}]{Chodos1980} Chodos A. and Detweiler S., 1980,
Phys. Rev. {\bf D21} 2167.

\bibitem[\protect\citeauthoryear{Dyer \& Marleau}{1995}]{Dyer1995} Dyer C. and Marleau F., 1995,
Phys. Rev. {\bf D13} 5588.

\bibitem[\protect\citeauthoryear{Einstein}{1911}]{Einstein1911} Einstein A., 1911, Ann. Phys. 35.

\bibitem[\protect\citeauthoryear{Einstein}{1915}]{Einstein1915} Einstein A., 1915, Kon. Preuss. Akad. Wissen.
(Berlin) Sitzungs. 112.

\bibitem[\protect\citeauthoryear{Eubanks et al.}{1999}]{Eubanks1999} Eubanks, T. M. et al., 1999,
Advances in solar system tests of gravity, [preprint]
ftp://casa.usno.navy.mil/navnet/postscript/, file prd 15.ps.

\bibitem[\protect\citeauthoryear{Fukui}{1987}]{Fukui1987} Fukui T., 1987, Gen. Rel. Grav. {\bf 19} 43.

\bibitem[\protect\citeauthoryear{Gegenberg \& Das}{1985}]{Gegenberg1985} Gegenberg J. D. and Das A., 1985,
Phys. Lett. {\bf A112} 427.

\bibitem[\protect\citeauthoryear{Iba`nez \& Verdaguer}{1986}]{Iba`nez1986} Iba`nez J. and Verdaguer E., 1986,
Phys. Rev. {\bf D34} 1202.

\bibitem[\protect\citeauthoryear{Ishihara}{1984}]{Ishihara1984} Ishihara H., 1984, Prog. Theor. Phys. {\bf 72} 376.

\bibitem[\protect\citeauthoryear{Iorio}{2005a}]{Iorio2005a} Iorio L., 2005a, JCAP {\bf 9} 6.

\bibitem[\protect\citeauthoryear{Iorio}{2005b}]{Iorio2005b} Iorio L., 2005b, arxiv:gr-qc/0511138.

\bibitem[\protect\citeauthoryear{Kagramanova, Kunz \& L{\"a}mmerzahl}{2006}]{Kagramanova2006} Kagramanova V.,
Kunz J. and L{\"a}mmerzahl C. L., 2006 [arxiv:gr-qc/0602002].

\bibitem[\protect\citeauthoryear{Kalligas, Wesson \& Everitt}{1995}]{Kalligas1995} Kalligas D., Wesson P. S.
and Everitt C. W. F., 1995, Astrophys. J. {\bf 439} 548.

\bibitem[\protect\citeauthoryear{Lebach et al.}{1995}]{Lebach1995} Lebach D. E. et al., 1995, Phys. Rev. Lett. {\bf
75} 1439.

\bibitem[\protect\citeauthoryear{Lim, Overduin \& Wesson}{1995}]{Lim1995} Lim P. H., Overduin J. M. and Wesson P. S.,
1995, J. Math. Phys. {\bf 36} 6907.

\bibitem[\protect\citeauthoryear{Liu \& Overduin}{2000}]{Liu2000} Liu H. and Overduin J. M., 2000,
Astrophys. J. {\bf 538} 386.

\bibitem[\protect\citeauthoryear{Mayers \& Perry}{1986}]{Mayers1986} Mayers R. and Perry M., 1986,
Annal. Phys. {\bf 172} 304.

\bibitem[\protect\citeauthoryear{Newton}{1704}]{Newton1704} Newton, I., "Opticks", Dover, 1979.

\bibitem[\protect\citeauthoryear{Nordtvedt}{2001}]{Nordtvedt2001} Nordtvedt K., 2001,
Phys. Rev. {\bf D61} 122001.

\bibitem[\protect\citeauthoryear{Ponce de Leon}{2003}]{Ponce2003} Ponce de Leon J., 2003,
Gen. Rel. Grav. {\bf 35} 1365.

\bibitem[\protect\citeauthoryear{Pound \& Rebka}{1959}]{Pound1959} Pound, R. V. and Rebka Jr. G. A., 1959,
Phys. Rev. Lett. {\bf 3} 439.

\bibitem[\protect\citeauthoryear{Pound \& Rebka}{1960}]{Pound1960} Pound R. V. and Rebka Jr. G. A., 1960,
Phys. Rev. Lett. {\bf 4} 337.

\bibitem[\protect\citeauthoryear{Pound \& Snider}{1964}]{Pound1964} Pound R. V. and Snider J. L., 1964,
Phys. Rev. Lett. {\bf 13} 539.

\bibitem[\protect\citeauthoryear{Rahaman et al.}{2005a}]{Rahaman2005a} Rahaman F. et al., 2005,
Mod. Phys. Lett. {\bf A20} 1627.

\bibitem[\protect\citeauthoryear{Rahaman et al.}{2005b}]{Rahaman2005b} Rahaman F. et al., 2005,
Int. J. Mod. Phys. {\bf A20} 993.

\bibitem[\protect\citeauthoryear{Ray}{2006}]{Ray2006} Ray S., 2006, Int. J. Mod. Phys. {\bf D15} 917.

\bibitem[\protect\citeauthoryear{Schwarz}{1985}]{Schwarz1985} Schwarz J. H., 1985,
Superstings, (World Scientific, Singapore).

\bibitem[\protect\citeauthoryear{Shapiro}{1964}]{Shapiro1964} Shapiro I. I., 1964, Phys. Rev. Lett. {\bf 13} 789.

\bibitem[\protect\citeauthoryear{Shapiro, Counselman \&
King}{1976}]{Shapiro1976} Shapiro I. I., Counselman C. C. and King
R. W., 1976, Phys. Rev. Lett. {\bf 36} 555.

\bibitem[\protect\citeauthoryear{Shapiro, Davis, Lebach \& Gregory}{2004}]{Shapiro2004} Shapiro S. S., Davis J. L.,
Lebach D. E. and Gregory J. S., 2004, Phys. Rev. Lett. {\bf 92} 121101.

\bibitem[\protect\citeauthoryear{Weinberg}{1986}]{Weinberg1986} Weinberg S., 1986, Strings and Superstrings,
(World Scientific, Singapore).

\bibitem[\protect\citeauthoryear{Weinberg}{2004}]{Weinberg2004} Weinberg S., 2004,
Gravitation and Cosmology, (Weilly Eastern, Inc. p. 203).

\bibitem[\protect\citeauthoryear{Wesson}{1983}]{Wesson1983} Wesson P. S., 1983, Astron. Astrophys. {\bf 119} 145.

\bibitem[\protect\citeauthoryear{Will}{2001}]{Will2001} Will C. M., 2001, arxiv: gr-qc/0103036.
\end{thebibliography}
 \end{document}